\pdfoutput=1
\documentclass[preprint,12pt,authoryear]{elsarticle}

\usepackage[utf8]{inputenc}
\usepackage[T1]{fontenc}
\usepackage{lmodern}
\usepackage{graphicx}
\usepackage{booktabs}
\usepackage{multirow}
\usepackage{amsmath,amssymb}
\usepackage{orcidlink}
\usepackage{hyperref}
\hypersetup{hidelinks}
\makeatletter
\def\ps@pprintTitle{%
  \let\@oddhead\@empty
  \let\@evenhead\@empty
  \def\@oddfoot{\footnotesize\itshape Preprint\hfill\today}%
  \let\@evenfoot\@oddfoot}
\makeatother

\begin{document}

\begin{frontmatter}

\title{Who Speaks for the Water? Simulating Discursive Closure and Power Asymmetries in Hydrosocial Governance}

\author[udec]{Mario Lillo-Saavedra\,\orcidlink{0000-0001-5634-9162}\corref{cor}}
\ead{malillo@udec.cl}
\author[udecplan]{Marcela Salgado-Vargas\,\orcidlink{0000-0002-1353-375X}}
\author[garcia]{\'Angel Garc\'ia-Pedrero\,\orcidlink{0000-0002-6848-481X}}
\author[udd]{Diego Rivera\,\orcidlink{0000-0003-0916-1540}}
\author[baylor]{Marcelo Somos-Valenzuela\,\orcidlink{0000-0001-7863-4407}}

\cortext[cor]{Corresponding author:}

\affiliation[udec]{organization={Departamento de Mecanizaci\'on y Energ\'ia, Facultad de Ingenier\'ia Agr\'icola, Universidad de Concepci\'on},
            city={Chill\'an},
            country={Chile}}
\affiliation[udecplan]{organization={Departamento de Planificaci\'on Territorial y Sistemas Urbanos, Facultad de Ciencias Ambientales, Universidad de Concepci\'on},
            city={Concepci\'on},
            country={Chile}}
\affiliation[garcia]{organization={Department of Computer Architecture and Technology, School of Computer Engineering, Universidad Polit\'ecnica de Madrid},
            city={Boadilla del Monte},
            country={Spain}}
\affiliation[udd]{organization={Centro de Sustentabilidad y Gesti\'on Estrat\'egica de Recursos (CiSGER), Facultad de Ingenier\'ia, Universidad del Desarrollo},
            country={Chile}}
\affiliation[baylor]{organization={Department of Geosciences, Baylor University},
            city={Waco, TX},
            country={USA}}

\begin{abstract}
Water governance conflicts are shaped not only by material scarcity 
but by power asymmetries, institutional distrust, and the discursive 
frameworks through which actors assign meaning to water, 
infrastructure, and legitimate authority. Conventional simulation 
tools represent these dynamics through numerical opinion variables, 
thereby missing the argumentative and narrative dimensions that 
structure real deliberation in hydrosocial territories. We present 
the Hydrosocial Agent-Based Model with Large Language Models 
(HSABM--LLM), a hybrid framework that enables heterogeneous agents 
to generate, evaluate, and adapt natural language arguments 
conditioned on institutional roles, interaction histories, and 
territorial context. Grounded in semi-structured interviews, a 
structured survey of water users ($n = 293$), and 456 press and 
official documents from the Longav\'i River sub-basin (Maule Region, 
Chile), the model simulates deliberation over a contested large-scale 
reservoir proposal across four experimental scenarios. Opinion 
convergence toward unanimous opposition emerged within 7--14 months, 
operating through discursive closure driven by distributive equity 
and water security framings rather than argumentative dominance. 
Governance actors accelerated convergence by activating pre-existing 
institutional distrust rather than through direct persuasion, while 
peer-based smallholder interactions accounted for 77--79\% of all 
opinion changes. The formal authority of state institutions proved 
structurally decoupled from deliberative influence, revealing a 
legitimacy deficit that formal inclusion alone cannot resolve. These 
findings suggest that participatory governance design must attend to 
the discursive conditions under which institutional actors enter 
deliberative spaces, rather than assuming that inclusion is 
inherently consensus-building. The HSABM--LLM framework offers a 
novel platform for integrating political ecology insights into 
computational social science.
\end{abstract}

\begin{keyword}
Hydrosocial systems \sep Discursive dynamics \sep Agent-based modeling \sep Large language models
\end{keyword}

\end{frontmatter}

\section{Introduction}

Hydrological dynamics, institutional arrangements, struggles over legitimacy, and power relations embedded in knowledge practices shape water crises beyond the gap between supply and demand \citep{Boelens2014, Swyngedouw2009, Zeitoun2020, ling2022global, quandt2022role, Salgado2025}. Hydrosocial territories thus emerge as socioecological assemblages rather than purely technical allocation systems. Under increasing water stress, technocratic instruments and fragmented regulations struggle to integrate diverse values, knowledge, and expectations \citep{Ormsbee2024Need,bauer2015water}. Water conflicts are therefore socio-environmental struggles within hydrosocial systems, where material, normative, and symbolic assemblages define access to water, allocation rules, and epistemic recognition for different actors.

Tension among stakeholders does not follow linear trajectories but rather evolves through cycles of escalation, negotiation, or reconfiguration that intersect with climatic variability, institutional weaknesses, and struggles over legitimacy \citep{zeitoun2020analysis, blanco2023water}. Power asymmetries shape how different actors interpret water-related problems, responsibilities, and claims. Moreover, climate change has increased hydrological variability and competition among uses \citep{WMO2024, onyena2025blue, yasmeen2024traits,rivera2026decadal}, intensifying pressures on an institutional system characterized by longstanding inequalities.

Discourse structures water conflicts, operating as broad normative frameworks that define which problems gain relevance, which solutions are considered legitimate, and which forms of knowledge achieve institutional visibility \citep{Boelens2014, Strang2023, De2024Water}. Actors produce situated narratives articulating positions, demands, and explanations regarding hydrosocial processes. Narratives of scarcity, efficiency, modernization, or risk express and mobilize these discursive frameworks, guiding public interventions and contributing to the legitimation of specific regulatory arrangements \citep{Zwarteveen2017, Zwarteveen2021, Hommes2023Need, 
Swyngedouw2009, Zeitoun2020}.

In Latin America, empirical studies on water narratives, such as participatory and decolonial approaches \citep{hasbun2021investigacion, VelezGaleano2018, Dupuits2023, Sullivan2024, Puppo2025}, have primarily relied on qualitative methods \citep{Boelens2008, Figueroa2020, LopezFlores2024}. Qualitative research has been essential for documenting the symbolic and political dimensions of water governance, but a downside is its labor-intensive nature and reliance on situated analysis, which limits its integration into formal hydrosocial models. However, existing simulation tools use deterministic architectures, fixed rules or aggregated variables, which constrain their capacity to represent contested meanings, narrative practices, or transformations in legitimacy \citep{Taberna2020, Lillo2024Socio}.

Large language models (LLMs), understood as systems trained on large textual corpora to generate and evaluate natural language, have been used to identify themes, interpretative frames, and affective content across large document collections \citep{larosa2025, Batista2025}. LLMs enable the synthesis of heterogeneous information and the detection of discursive patterns; however, they require human oversight due to response variability, biases, and limitations in contextual interpretation. LLMs can recognize narrative dimensions such as temporality, focalization, and agency, but they struggle to identify conflict, abstraction, and experiential depth \citep{PiperBagga2024NarrativeLLM}. Because LLMs are trained on multilingual and multicultural corpora, they can nonetheless capture discursive nuances such as irony, emotional appeals, and local references \citep{masoud2024, florez2025latamgpt}. Still, they tend to produce simplified arcs and less diverse structures than those observed in human storytelling \citep{Tian2024HumanNarratives}, which constrains their direct application to hydrosocial modeling.

Agent-based modeling (ABM) addresses some limitations by incorporating heterogeneity, interaction, and negotiation \citep{Blair2022, Alam2022Understanding, Sousa2024, Lillo2024Socio}, although many models retain assumptions of instrumental rationality and fixed rules that hinder the representation of narrative dynamics, customary norms, or shifts in legitimacy \citep{Bourceret2021Governance, Taberna2020}. Recent hybrid architectures combining ABM and LLMs revisit constraints by enabling generative agents to interpret context and produce arguments through natural language \citep{manduchi2025, thapa2025large, adornetto2025generative, lu2024llms}, while introducing new challenges related to semantic consistency, narrative drift, and adaptation to specific cultural contexts \citep{Ghaffarzadegan_2024, dejeu2024}.

The Chilean 1981 Water Code consolidated a market-oriented governance regime that favored actors with legal and economic capacity, while smallholder farmers, indigenous people, and environmental organizations occupied subordinated positions \citep{bauer2015water, budds2013water, budds2004power}. We use as a case study a proposal to build a large-scale reservoir in the Longaví River sub-basin \citep{Salgado2025} that, with the water code as a governance framework, becomes a hydrosocial territory. We implemented a hydrosocial agent-based model with LLM-based reasoning (HSABM--LLM) to represent discursive dynamics. The model couples a monthly socio-hydrological ABM with a language module that generates, evaluates, and adapts arguments based on institutional rules, access to information, and prior actor interactions. In the model, narratives are evolving trajectories shaped by unequal capacities to participate in and influence water governance debates, rather than static attributes. Thus, the objectives are: (i) to integrate narratives and power relations into hydrosocial modeling through a hybrid ABM--LLM architecture, and (ii) to simulate how institutional and interactional configurations influence the acceptance, contestation, or reconfiguration of water interventions, such as large infrastructure. Our results link qualitative approaches to conflict and simulation-based tools, while the case study identifies direct implications for adaptive governance in asymmetric hydrosocial contexts.

\section{Methodology}

\subsection{Study Area}

\begin{figure}[ht]
\centering
\includegraphics[scale=0.3]{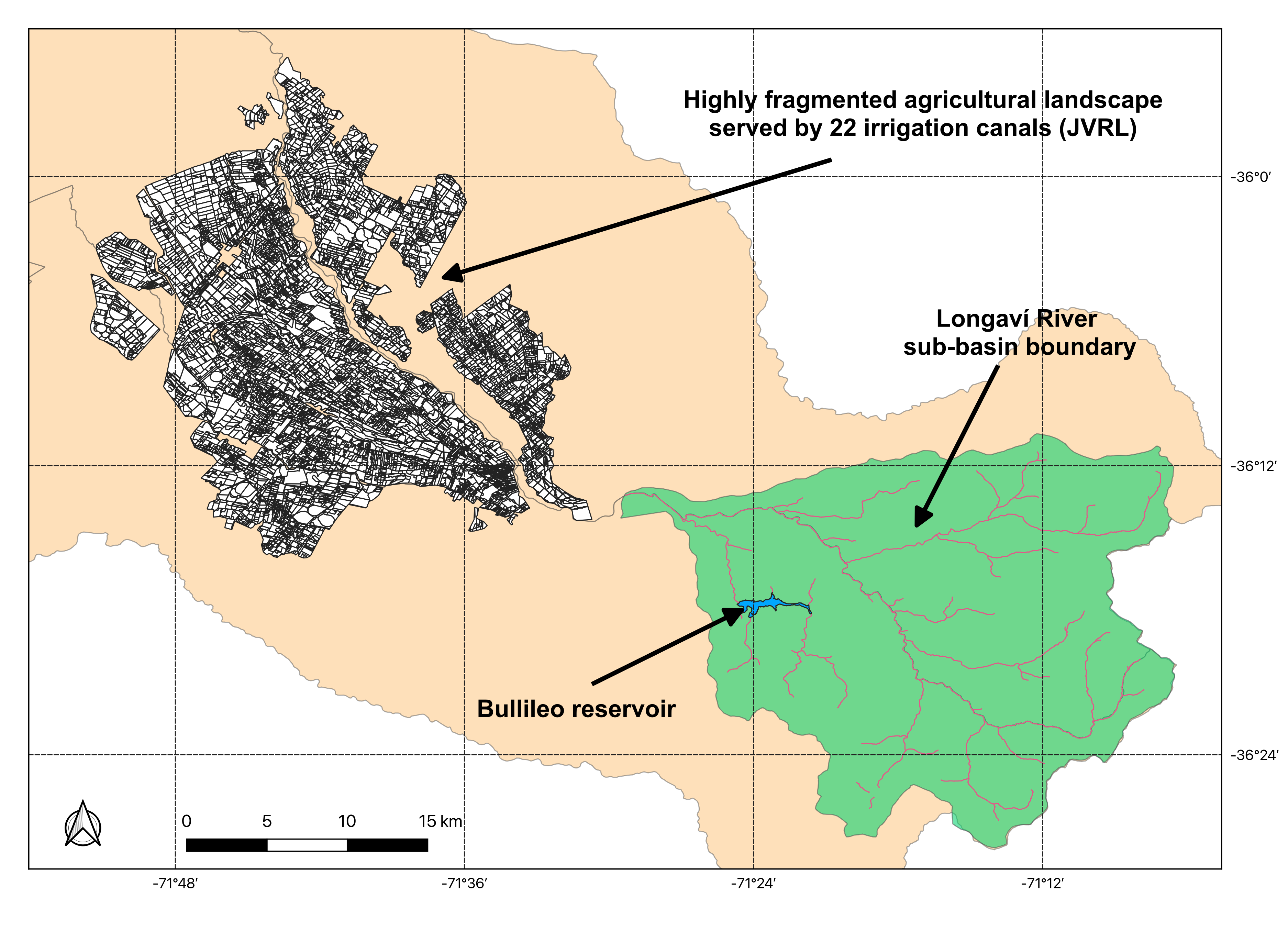}
\caption{Study area in the Longaví River basin, Maule Region, Chile
($36^\circ18'$\,S, $71^\circ18'$\,W). Black polygons show the highly
fragmented agricultural landscape under the Junta de Vigilancia del
Río Longaví (JVRL) jurisdiction, served by 22 irrigation canals. The
Longaví River sub-basin (676\,km$^2$, green) features a pluvio-nival
regime, with the Bullileo reservoir (60\,Mm$^3$) providing seasonal
storage to offset the mismatch between winter precipitation and summer
irrigation demand.}
\label{fig:mapa}
\end{figure}

The Longaví River sub-basin, Maule Region, Chile ($36^\circ18'~S$, 
$71^\circ18'~W$), covers $676~\text{km}^2$ (Figure \ref{fig:mapa}). 
Land use is a highly fragmented agricultural landscape (nearly 80\,\% 
of farmers hold less than $12~\text{ha}$) interspersed with forestry. 
Water infrastructure comprises a reservoir of $60~\text{Mm}^3$ and 22 
irrigation canals. The climate is Mediterranean, with a pluvio-nival 
streamflow regime. Over 75\,\% of annual precipitation occurs during 
austral fall-winter (May--August), while maximum irrigation demand 
peaks between September and March \citep{CR2}. Thus, the seasonal 
mismatch between water availability and demand requires water storage.
The Junta de Vigilancia del Río Longaví (JVRL) coordinates water 
governance within the sub-basin. As a private, non-profit water users' 
organization regulated under the Chilean Water Code, its primary 
mandate is to manage and allocate water rights among its members in 
natural watercourses.
The sub-basin constitutes a hydrosocial territory 
marked by power asymmetries and distributional disputes, and the 
fieldwork conducted by \citet{Salgado2025} provides the empirical 
foundation for the model developed in this study.

\subsection{HSABM--LLM Modeling Approach}

The HSABM--LLM model represents hydrosocial dynamics, discursive practices, and power relations by coupling an agent-based model at monthly time steps. The LLM module generates, evaluates, and adapts arguments in response to changes in available information, institutional rules, and actor interactions.

The design rests on two empirical assumptions \citep{smith2021nlp, kosovac2024discourses, biancardi2022}: (i) discourses on water sustainability exhibit systematic configurations amenable to semantic representation; and (ii) interactions are mediated by asymmetric capacities to produce and circulate legitimate narratives. Operationally, assumptions translate into heterogeneous agent profiles with differentiated discursive influence and institutional weight, LLM-generated discursive states at each time step, and explicit communication rounds incorporating interaction memory and an evaluator module. The structure follows previous findings on cooperation in multi-agent systems with LLM-based reasoning \citep{NeurIPS2024}.

\subsection{Empirical Grounding and Data Sources}
\label{subsec:parametrizacion}

The HSABM--LLM configuration draws on a mixed empirical corpus that grounds agent behavior in the observed dynamics of the Longaví River sub-basin. Parameterization integrates the qualitative work of \citet{Salgado2025}, comprising nine semi-structured interviews, field notes, and a structured survey of JVRL water users ($n=293$, August 2024). Agent profiles were derived analytically from survey results, capturing recurrent configurations of vulnerability, organizational embeddedness, and institutional distrust, rather than through purely statistical clustering.

Discursive dynamics were categorized through inductive coding in Atlas.ti from open-ended survey responses and interviews, yielding a taxonomy of discursive categories organized around three dimensions: conflict types (e.g., obstruction, water diversion), attribution of responsibility (e.g., perceived state abandonment), and preferred solution strategies. Coding continued iteratively until theoretical saturation, i.e., no new categories or conceptual dimensions emerged from additional data across the three source types. We also used a review of 456 press articles and official documents, from which the argument matrix used by agents in the simulation was 
constructed.

\subsection{Computational Simulation Framework}
\label{sec:simulation_framework}
The HSABM--LLM simulator was implemented in Python using a custom architecture, without pre-built simulation libraries, to maximize state management efficiency and synchronization with the inference module. The cognitive component uses Qwen3-8B, a causal 
language model \citep{radford2018improving, qwenteam2025qwen3} with approximately 8.2 billion parameters, 36 transformer layers, and GQA attention (32 Q-heads, 8 KV-heads). The setup enables agents to alternate between structured logical inference and natural role-playing dialog, maintaining profile coherence in Spanish without the computational overhead of larger models. Simulations run locally on an Apple MacBook Pro (M2 Pro, 16 GB RAM) with Metal Performance Shaders (MPS) hardware acceleration.

The framework models water conflict dynamics through $N$ 
heterogeneous agents interacting over a dynamic social network 
(Eq.~\ref{eq:grafo}):
\begin{equation}
G^{(t)} = (V, E^{(t)})
\label{eq:grafo}
\end{equation}
where $V$ is the static set of actors and $E^{(t)}$ the active 
communication links at discrete time step $t = 1, \dots, T$. 
Interactions are restricted to valid edges $(i, j) \in E^{(t)}$, 
whose activation probability depends on an influence matrix and 
functional hierarchies derived from fieldwork, including 
semi-structured interviews, survey responses, and press document 
analysis (Section~\ref{subsec:parametrizacion}). Additionally, 
interaction intensity, defined as the fixed proportion of active 
links per cycle, must be specified, as it 
determines how many of the valid edges are activated at each 
time step and is treated as an experimental factor in the 
simulation design.

In this hybrid architecture, the ABM serves as the structural orchestrator, while the LLM handles the cognitive and semantic dimensions. The ABM performs three critical functions: (i) \textit{scheduling}, controlling agent activation sequences at each step $t$ to synchronize hydrosocial events with negotiation cycles; (ii) state management, tracking non-discursive variables $\mathbf{x}_i$ and external constraints that feed LLM prompts; and (iii) relational dynamics, updating network topology $G^{(t)}$ and selecting interaction pairs.

Each agent $i \in \{1, \dots, N\}$ is defined by a state tuple $A_i = (\mathbf{x}_i,\, o_i^{(t)},\, h_i^{(t)})$, where $\mathbf{x}_i$ is the vector of static socio-structural attributes (e.g., farm size, irrigation type, institutional role) derived from Section~\ref{subsec:parametrizacion}; $o_i^{(t)} \in [-1, 1]$ is the agent's stance toward the evaluated proposal; and $h_i^{(t)}$ is the accumulated discursive history. Contextual prompts incorporating $\mathbf{x}_i$ and the active scenario description generate the initial opinion $o_i^{(0)}$ and the first narrative argument.

Agent inference relies on dynamic prompt templates injecting three levels of context: agent identity (from $\mathbf{x}_i$), current system state (narrative summary $S^{(t)}$), and receiver profile $j$. Persuasive message generation is (Eq.~\ref{eq:mensaje_per}):
\begin{equation}
m_{i \to j}^{(t)} \sim f_{\text{LLM}}
(\mathcal{P}(\mathbf{x}_i, \mathbf{x}_j, S^{(t)}, 
h_i^{(t-1)}); \tau)
\label{eq:mensaje_per}
\end{equation}
where $\mathcal{P}$ constructs the prompt from sender and receiver attributes, a narrative summary, and prior discursive history. The temperature parameter $\tau$ controls the stochasticity of the language model's output distribution: lower values yield more deterministic responses, while higher values increase diversity at the cost of coherence. 

Following \citet{renze2024effect} and calibration tests, $\tau = 0.7$ aims to balance structural coherence and narrative diversity, avoiding mechanical repetition 
($\tau < 0.5$) and role inconsistency ($\tau \approx 1$).

For each active pair $(i,j)$, the sender generates message $m_{i \to j}^{(t)}$ and the receiver produces an argumentative response $r_{j \to i}^{(t)}$ (acceptance, rejection, or 
counter-argument). The independent evaluator module infers positional shifts as in Eq.~\ref{eq:regla_actua}:
\begin{equation}
o_j^{(t)} = o_j^{(t-1)} + \alpha_j \cdot 
\Delta o(r_{j \to i}^{(t)})
\label{eq:regla_actua}
\end{equation}
where $\Delta o \in \{-1, 0, 1\}$ is the inferred direction of change and $\alpha_j \in [0, 1]$ is agent $j$'s intrinsic susceptibility, conditioned by its functional profile. Polarization or consensus thus emerges as a system-level property driven by argumentative resonance rather than network pressure alone.

At the end of each step $t$, all exchanged messages $\mathcal{M}^{(t)}$ are aggregated into an updated narrative summary $S^{(t)} = f_{\text{sum}}(\mathcal{M}^{(t)} \cup S^{(t-1)})$, which $f_{\text{sum}}$ identifies dominant topics, emerging conflicts, and shifts in discourse. This summary is injected as shared contextual memory into the prompts for cycle $t+1$, enabling agents to adapt their discursive strategies to emergent system-level trends and to couple individual behavior to collective narrative dynamics.

\subsection{Simulation Scenario and Experimental Design}

The model simulates the evaluation phase of a large-scale multipurpose reservoir (80 Mm$^3$) in the Longaví River sub-basin. From fieldwork, the reservoir is a focal point of conflict and deliberation, concentrating expectations of water security alongside concerns over distributive and environmental impacts; therefore, it is perceived as a high socio-environmental impact intervention \citep{Salgado2025}. Water supply equals average flow conditions, excluding extreme drought; thus, scarcity is not the primary driver of decisions. 

 Water uses encompass agriculture, human needs, agro-industry, and tourism. The Dirección de Obras Hidráulicas (DOH), the Chilean government agency responsible for planning and executing hydraulic infrastructure, manages the project. Fieldwork revealed moderate trust in the JVRL, low trust in the DOH, unequal organizational participation, and a low initial conflict level with escalation potential driven by perceived distributive inequity or environmental harm \citep[cf.][]{Salgado2025}. 

Water availability in the Longaví River sub-basin follows a sequential 
distribution from headwaters to canal tail-ends, with 
progressive losses and extractions along the network. 
Consequently, an agent's longitudinal position within the 
canal system conditions its effective water access and, 
by extension, its vulnerability to distributive impacts 
of the proposed reservoir, shaping the initial discursive 
dispositions incorporated into agent profiles.

Experiments analyze the evolution of the discursive trajectory 
over a $T = 24$-month horizon with monthly time steps. Initial 
opinions are Support ($o_i^0 = 1$), Opposition ($o_i^0 = -1$), 
or Neutrality ($o_i^0 = 0$), distributed to yield a heterogeneous 
discursive field without a dominant stance. The simulated 
population comprises seven fixed agent types: \textit{smallholder 
farmers}, small-scale producers ($\leq$12~ha) with high 
agricultural dependency, limited institutional trust, and 
vulnerability to positional disadvantage within the canal 
network; \textit{technified producers}, medium-to-large operators 
with irrigation investment and technical orientation, open to the 
reservoir but demanding clear governance rules; \textit{canal 
presidents}, elected representatives of water user communities 
mediating between smallholders and institutions with 
equity-centered mandates; a \textit{forestry company}, a large 
territorial and hydraulic actor (3,086~ha; $>$1,700~l/s), 
favorable to the reservoir but skeptical of expanded state 
governance; the \textit{DOH}, a state hydraulic agency acting 
as a non-originating node that collects and systematizes agent 
arguments without persuasive capacity; an \textit{environmental 
NGO}, ecosystem-oriented and conditionally critical of large 
infrastructure, focused on equity and ecological impact framings; 
and the \textit{JVRL}, a second-level water user organization 
with legal mandate over basin governance, proactive in 
infrastructure planning but operating under a documented 
legitimacy deficit among users.

\begin{table}[ht]
    \centering
    \caption{Experimental design: varied factors and fixed population parameters across scenarios, including initial opinion distribution ($F$: Pro, $C$: Con, $N$: Neutral).}
    \label{tab:diseno_experimentos}
    \begin{tabular}{lcccc}
        \hline
        & \multicolumn{4}{c}{\textbf{Scenario}} \\
        \cline{2-5}
        & \textbf{ST02} & \textbf{ST09} & \textbf{TT02} & \textbf{TT09} \\
        \hline
        \multicolumn{5}{l}{\textit{Experimental factors}} \\
        \hline
        JVRL and NGO included   & No  & No  & Yes & Yes \\
        Interaction probability & 0.2 & 0.9 & 0.2 & 0.9 \\
        \hline
        \multicolumn{5}{l}{\textit{Fixed population parameters (Initial option $F/C/N$)}} \\
        \hline
        Smallholder farmers       & \multicolumn{4}{c}{75 (30/30/15)} \\
        Technified producers       & \multicolumn{4}{c}{25 (10/10/5)} \\
        Canal presidents            & \multicolumn{4}{c}{3 (1/1/1) }  \\
        Forestry company            & \multicolumn{4}{c}{1 (1/0/0) }  \\
        JVRL representative          & \multicolumn{4}{c}{1 (1/0/0)}  \\
        DOH representative           & \multicolumn{4}{c}{1 (0/0/1)}  \\
        \hline
    \end{tabular}
\end{table}

For discursive trajectories, we classify each agent-generated argument into one of the nine semantic categories in the taxonomy (Table~\ref{tab:categorias_semanticas}), thus representing the dominant interpretive frameworks. Semantic categories came from inductive coding of the empirical corpus (interviews, open-ended survey responses, and press articles).

\begin{table}[ht]
    \centering
    \caption{Semantic categories used to classify agent arguments.}
    \label{tab:categorias_semanticas}
    \begin{tabular}{ll}
        \hline
        Code & Description \\
        \hline
        ECO  & Economic benefits and productivity gains \\
        SEG  & Water security and irrigation stability \\
        EQD  & Distributive equity in water access \\
        SOC  & Social capital and collective organization \\
        AMB  & Environmental and biodiversity impacts \\
        TERR & Territorial impacts on communities and land use \\
        POL  & Political pressure, subsidies, and state decisions \\
        ESC  & Skepticism, distrust, and uncertainty \\
        NEU  & Neutral or ambiguous arguments \\
        \hline
    \end{tabular}
\end{table}

\subsection{Aggregate Indicators}
\label{subsec:indicators}

To characterize system-level dynamics at each time step $t$, we define five aggregate indicators that summarize the simulation outputs analysed in the Results section. (i) The \textit{support level} measures the percentage of agents in favour of the reservoir proposal at time $t$, capturing the global trajectory of public opinion. (ii) \textit{Discursive diversity} is computed as the complement of the Simpson concentration index \citep{Simpson1949} over the $K = 9$ semantic categories (Table~\ref{tab:categorias_semanticas}), with higher values indicating a more pluralistic distribution of argumentative frames. (iii) The \textit{opinion change rate} (OCR) is the proportion of dyadic interactions that produce a shift in the receiver's stance, used as a proxy for deliberative permeability. (iv) The \textit{persuasion success rate by role} ($\text{PSR}_r$) records the share of interactions initiated by agents of role $r$ that result in an opinion change, allowing comparison of relative persuasive influence across actor types. (v) The \textit{semantic category frequency} ($f_k$) is the relative frequency of each semantic category $c_k$ across the full simulation horizon, describing the substantive composition of the deliberative field. Formal mathematical definitions of all five indicators are provided in Appendix~\ref{app:indicators}.

\section{Results}

This section presents the outcomes of the four simulation scenarios over 24 months, organized around aggregate opinion trajectories (Section~\ref{subsec:trayectorias}), the semantic composition of discursive dynamics (Section~\ref{subsec:semantica}), and interaction-level persuasion patterns (Section~\ref{subsec:persuasion}). Table~\ref{tab:resumen_resultados} provides a comparative summary of the main simulation indicators.

\begin{table}[ht]
    \centering
    \caption{Comparative summary of simulation outcomes across the four experimental scenarios. EQD is Distributive equity in water access (Table~\ref{tab:categorias_semanticas})}
    \label{tab:resumen_resultados}
    \begin{tabular}{lcccc}
        \hline
        & \multicolumn{4}{c}{\textbf{Scenario}} \\
        \cline{2-5}
        \textbf{Metric} & \textbf{ST02} & \textbf{ST09} & 
        \textbf{TT02} & \textbf{TT09} \\
        \hline
        Initial support (\%)        & 41 & 41 & 41 & 42 \\
        Peak support (\%)           & 52 & 46 & 41 & 42 \\
        Peak support step           & 2     & 6     & 0     & 0     \\
        Step at support $<10\%$     & 11    & 9     & 5     & 6     \\
        Step at support $= 0\%$     & 14    & 10    & 8     & 7     \\
        Total interactions          & 699   & 854   & 378   & 416   \\
        Opinion change rate (\%)    & 27  & 34  & 25  & 21  \\
        Final diversity (Simpson)   & 84 & 84 & 85 & 85 \\
        Dominant semantic category  & EQD   & EQD   & EQD   & EQD   \\
        \hline
    \end{tabular}
\end{table}

\subsection{Opinion Trajectories and Convergence Dynamics}
\label{subsec:trayectorias}

\begin{figure}[bt]
\centering
\includegraphics[width=\textwidth]{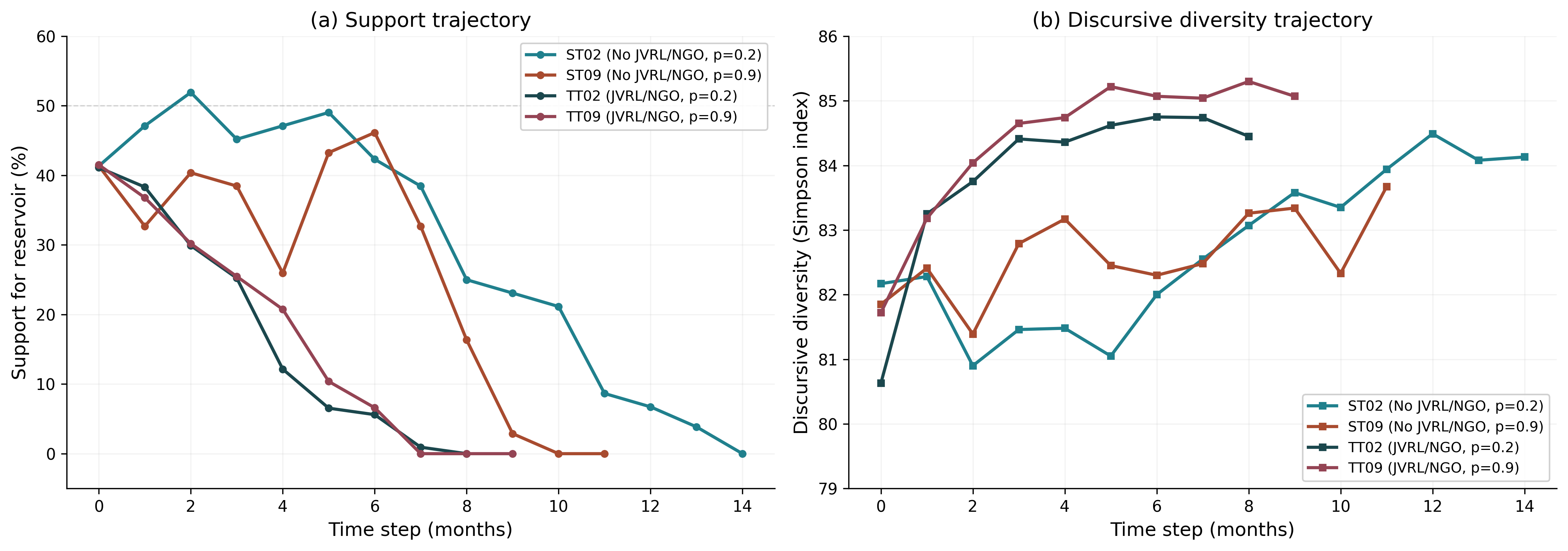}
\caption{(a) Support for the reservoir proposal over the simulation horizon across the four scenarios. (b) Discursive diversity measured by the Simpson index. Scenarios including the JVRL and NGO (TT02, TT09) exhibit a monotonic decline in support and higher terminal diversity.}
\label{fig:trayectorias}
\end{figure}

All four scenarios converged toward unanimous opposition ($o_i = -1$ for all agents) within the simulated 24-month period, although the trajectories exhibited distinct temporal profiles depending on institutional inclusion and interaction intensity (Figure~\ref{fig:trayectorias}). In scenarios without the JVRL and NGO agents (ST02 and ST09), support initially fluctuated before declining. In ST02 (low interaction probability, $p = 0.2$), support rose from 41\% to a peak of 52\% at step $t = 2$ months before declining monotonically, reaching zero at $t = 14$ months. The transient increase reflects early persuasive exchanges among supporters that could reinforce favorable stances before opposition arguments gained discursive traction. ST09 ($p = 0.9$) displayed a variable trajectory, oscillating between 26\% and 46\% through $t = 6$ months before a rapid collapse to zero by $t = 10$ months. High interaction density amplified persuasion in both directions in the early phases, generating a contested discursive space that eventually resolved through the accumulation of opposition arguments across a larger number of interaction pairs.

In scenarios with the JVRL and NGO (TT02 and TT09), support declined from the outset without exhibiting transient peaks. TT02 ($p = 0.2$) decreased from 41\% to zero at $t = 8$ months, while TT09 ($p = 0.9$) followed a similar trajectory, reaching zero at $t = 7$ months.  These figures suggest that the simulation captures a relational baseline in which the JVRL and allied organizations operate under a legitimacy deficit; consequently, their participation in the deliberative arena does not introduce neutral information but rather activates latent skepticism among other agents. The NGO exhibited a persuasion success rate exceeding 50\% in TT02 (Table~\ref{tab:persuasion_roles}), indicating that its framing of environmental and equity concerns resonated strongly with receiving agents, likely amplified by the pre-existing erosion of trust in formal water institutions.

Convergence speed thus varied across scenarios: the TT configurations reached unanimous opposition in 7--8 steps, whereas the ST configurations required 10--14 steps. This difference indicates that the inclusion of governance and environmental actors accelerated opinion consolidation by reducing the argumentative space available for reservoir supporters' positions, rather than merely adding more opponents to the population.

Discursive diversity, measured through the Simpson index (Figure~\ref{fig:trayectorias}b), showed a modest upward trend in all scenarios, rising from approximately 80--82 to 83--85 by the end of the simulation. This increase indicates that even as opinions converged toward opposition, agents continued to employ a diversifying repertoire of semantic categories to justify their positions. The TT scenarios yielded slightly higher terminal diversity values (85 in TT02 and TT09), consistent with the NGO and JVRL's introduction of environmental (AMB) and social capital (SOC) arguments, which expanded the discursive space beyond the economic and security framings that dominated the ST scenarios.

\subsection{Semantic Composition of Discursive Dynamics}
\label{subsec:semantica}

\begin{figure}[ht]
\centering
\includegraphics[width=\textwidth]{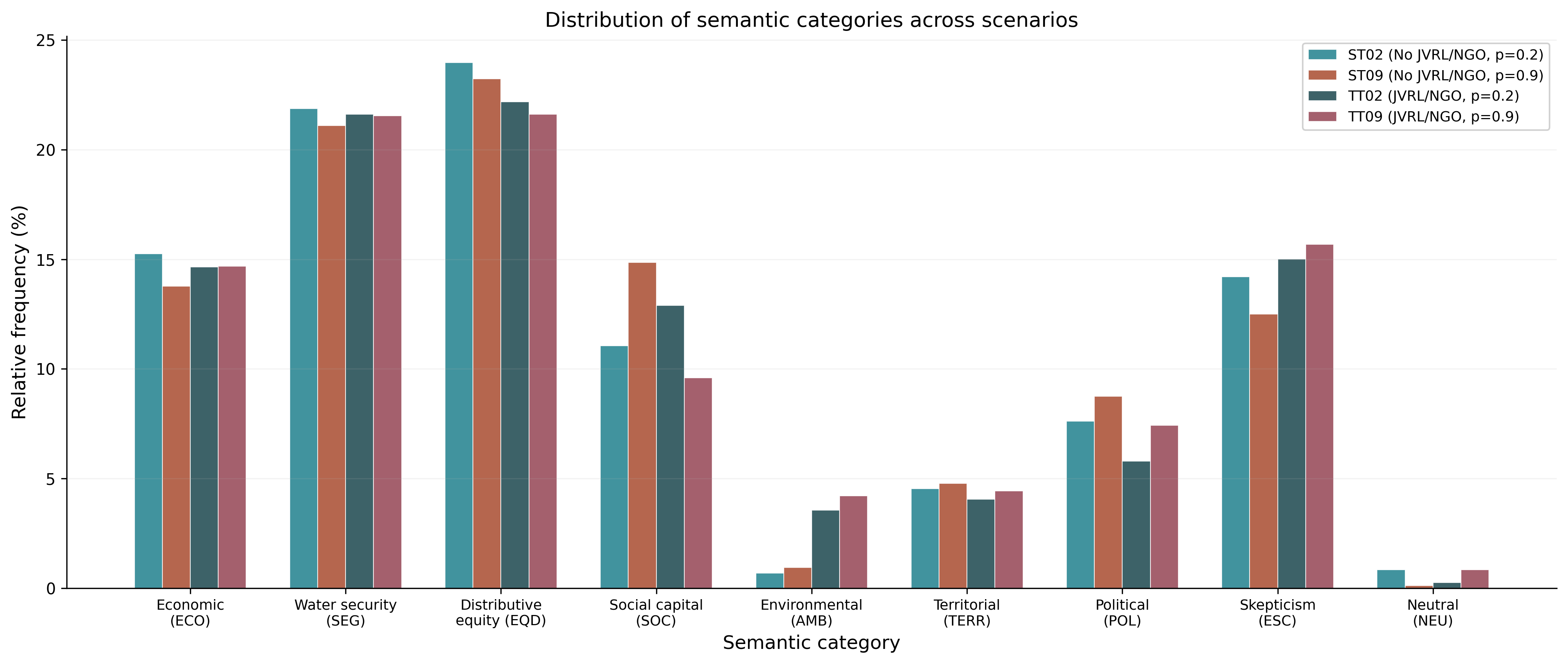}
\caption{Relative frequency of semantic categories across the four simulation scenarios. EQD, SEG, and ECO dominate across all configurations, while AMB increases with the inclusion of institutional actors (TT02, TT09).}
\label{fig:semantic_categories}
\end{figure}

The classification of agent-generated arguments into nine semantic categories shows a stable hierarchy across all scenarios. Distributive equity (EQD), water security (SEG), and economic benefits (ECO) are consistently the three most frequent positions (Figure~\ref{fig:semantic_categories}). EQD accounted for 21--24\% of all coded arguments, SEG for 21--22\%, and ECO for 14--15\%, reflecting the centrality of access, distribution, and livelihood concerns in the simulated deliberation process. However, systematic differences emerged across experimental conditions. Social capital (SOC) arguments are more frequent in ST09 (15\%) and TT02 (13\%) than in ST02 (11\%) or TT09 (10\%), suggesting that both high interaction density and the presence of governance actors stimulated references to collective organization and community-level coordination, though through distinct pathways. Skepticism (ESC) was elevated across all scenarios (12--16\%), with TT09 registering the highest relative frequency (16\%), indicating that institutional inclusion combined with high interaction intensity amplified expressions of distrust and uncertainty rather than resolving them.

Environmental arguments (AMB) remained marginal across all scenarios ($< 5\%$), but increased in the TT configurations: from 1\% (ST02 and ST09) to 4\% (TT02 and TT09). This increase is attributable to the NGO, which introduced biodiversity and ecological impact framings that were absent from the discursive repertoire of agricultural and institutional agents.

\begin{figure}[t]
\centering
\includegraphics[width=\textwidth]{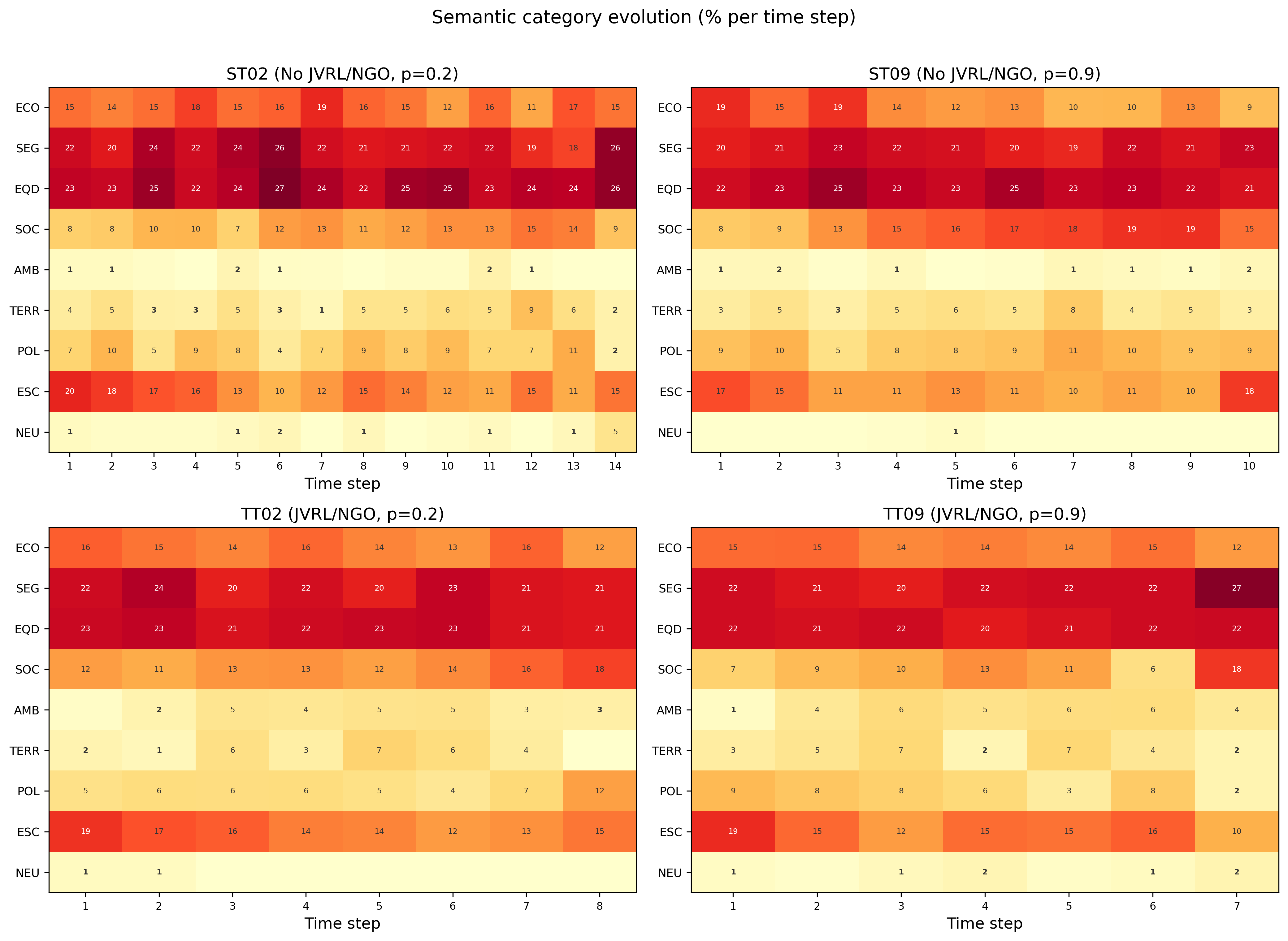}
\caption{Temporal evolution of semantic category proportions (\% per time step) across the four scenarios. EQD and SEG remain dominant throughout; AMB increases in TT scenarios; POL rises in terminal phases.}
\label{fig:topic_heatmaps}
\end{figure}

The temporal evolution of semantic categories (Figure~\ref{fig:topic_heatmaps}) shows that EQD and SEG maintained high and stable proportions throughout all scenarios, becoming structuring categories of the hydrosocial debate. Political pressure arguments (POL) showed an upward trend in the final steps of ST02 (rising from 7\% to 11\% at $t = 12$ months) and TT09, coinciding with phases in which most agents had already consolidated opposition and began articulating demands for institutional reform. The emergence of POL in the terminal phases suggests that, once positional consensus emerges, discursive dynamics shift from persuasion-oriented arguments to systemic claims about governance structures.

\subsection{Interaction Dynamics and Persuasion Patterns}
\label{subsec:persuasion}

\begin{figure}[ht]
\centering
\includegraphics[width=\textwidth]{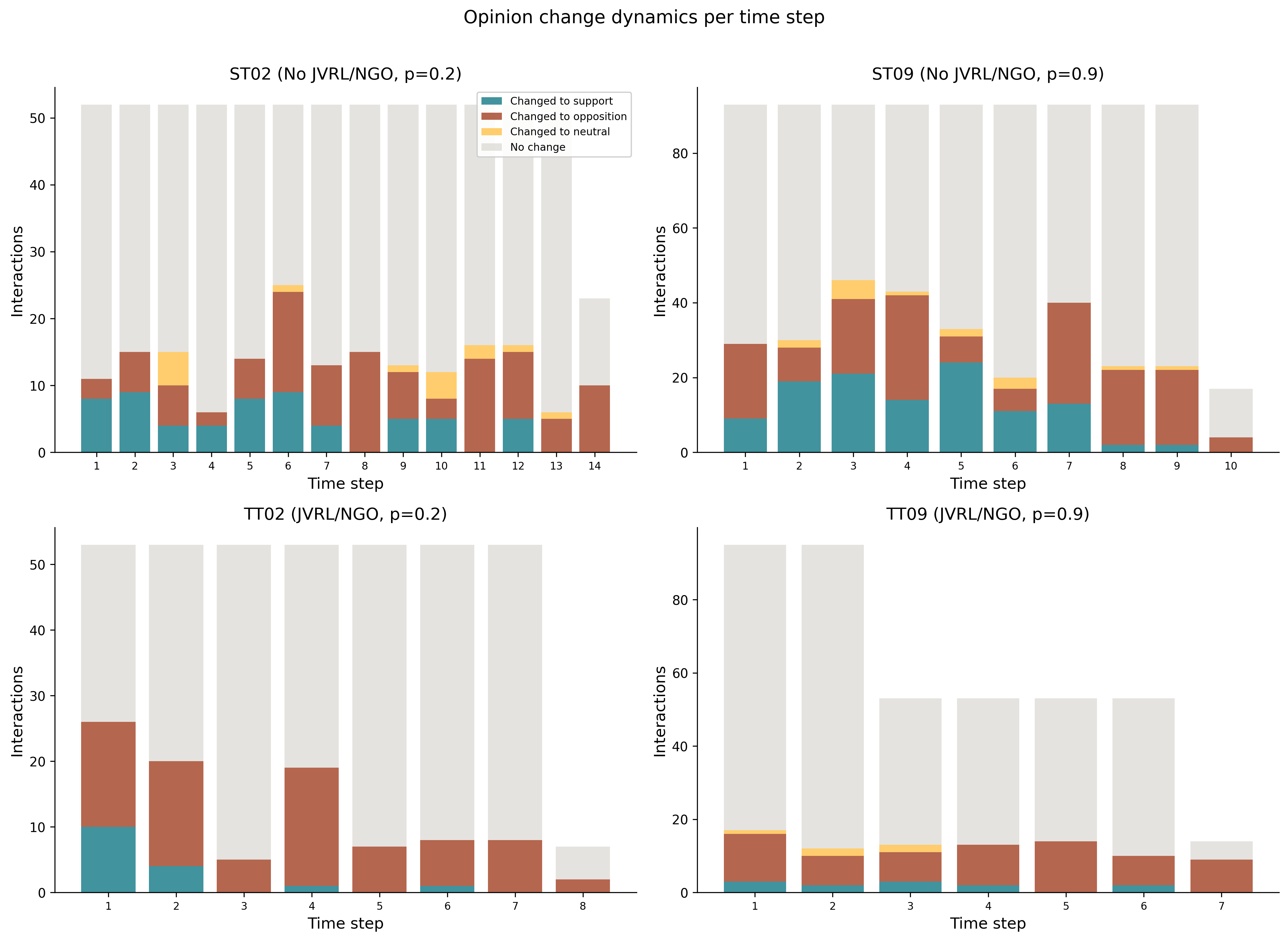}
\caption{Opinion change dynamics per time step across the four scenarios. Stacked bars indicate the number of interactions resulting in changes toward support (teal), opposition (terra), neutrality (gold), or no change (gray).}
\label{fig:opinion_dynamics}
\end{figure}

The stacked composition of opinion changes over time (Figure~\ref{fig:opinion_dynamics}) shows qualitative differences in how persuasion unfolds across scenarios. In ST02 and ST09, both directions of change (toward support and toward opposition) coexisted during the first half of the simulation, generating a contested deliberative space. In ST02, 1--6 months exhibited bidirectional changes with comparable magnitudes (e.g., month 6: 9 changes toward support vs. 15 toward opposition), whereas from 8 months onward, changes became exclusively toward opposition. In ST09, the higher interaction probability ($p = 0.9$) produced more volatile early dynamics, with month 3 registering 21 changes toward support and 20 toward opposition simultaneously, reflecting a highly contested discursive moment.

In the TT scenarios, opposition-directed changes dominated from 
the beginning. TT02 registered 16 changes toward opposition 
versus 10 toward support at $t = 1$~month, and by $t = 3$~months, 
changes toward support ceased. TT09 exhibited a similar pattern 
with 13 changes toward opposition versus 3 toward support at 
$t = 1$~month. This asymmetry indicates that the inclusion of 
institutional actors did not merely add participants but reshaped 
the argumentative structure of early interactions, establishing 
an opposition-favorable discursive environment from the outset.

\begin{table}[ht]
    \centering
    \caption{Persuasion effectiveness by sender role: 
    proportion of interactions resulting in opinion change 
    across the four scenarios.}
    \label{tab:persuasion_roles}
    \begin{tabular}{lcccc}
        \hline
        & \multicolumn{4}{c}{\textbf{Opinion change rate (\%)}} \\
        \cline{2-5}
        \textbf{Sender role} & \textbf{ST02} & \textbf{ST09} & 
        \textbf{TT02} & \textbf{TT09} \\
        \hline
        Smallholder farmer    & 29  & 36  & 32  & 27  \\
        Technified producer   & 19  & 30  & 11  & 6   \\
        Canal president       & 11  & 19  & 33  & 0   \\
        Forestry company      & ---   & ---   & ---   & ---   \\
        DOH                   & 0   & 0   & 0   & 10  \\
        JVRL                  & ---   & ---   & 0   & 0   \\
        NGO     & ---   & ---   & 67  & 25  \\
        \hline
    \end{tabular}
\end{table}

\begin{figure}[ht]
\centering
\includegraphics[width=\textwidth]{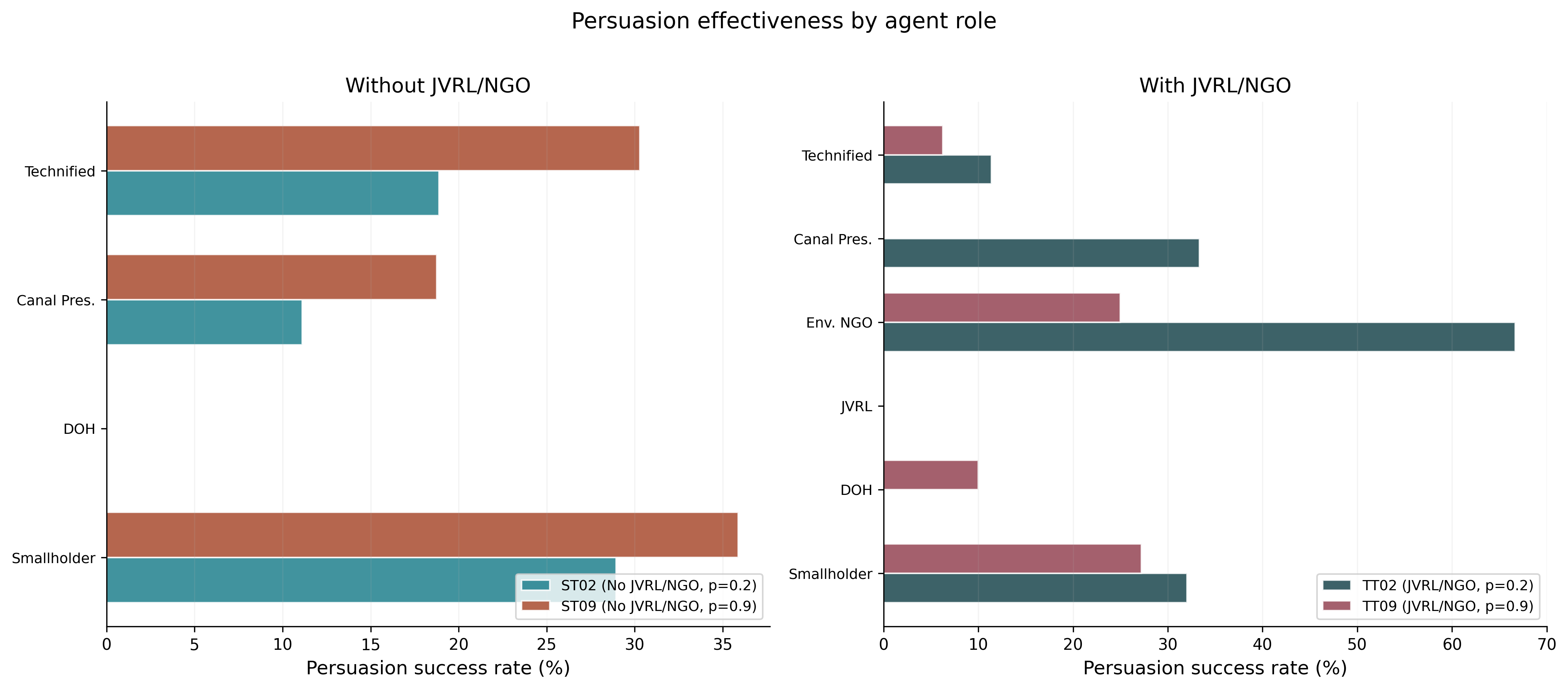}
\caption{Persuasion success rate by agent role, comparing scenarios without (left) and with (right) the JVRL and NGO. The NGO and canal presidents exhibit the highest per-interaction persuasion rates in the TT scenarios.}
\label{fig:persuasion_roles}
\end{figure}

At the role level (Table~\ref{tab:persuasion_roles} and Figure~\ref{fig:persuasion_roles}), smallholder farmers exhibited the highest aggregate persuasion effectiveness across most scenarios, benefiting from size dominance (75 agents) and the frequency of intra-group interactions. In ST09, the high interaction probability increased persuasion rates for all roles, with smallholders reaching 35.9\% and technified producers 30.3\%. The NGO achieved the highest per-interaction success rate in TT02 (66.7\%, 2 of 3 interactions) and TT09 (25.0\%, 1 of 4 interactions). However, these rates must be interpreted cautiously, given the small sample size inherent to a single-agent role.

The DOH, despite its institutional authority, registered a persuasion success rate of 0\% in three of the four scenarios and only 10\% in TT09. This finding indicates that the technocratic framing employed by the DOH, centered on aggregate efficiency and infrastructure optimization, failed to resonate with agents whose concerns were articulated through equity, autonomy, and distrust frames. The JVRL similarly registered no successful persuasion events in either TT scenario, suggesting that the coordination role within the simulation did not translate into discursive influence over individual agents' positions.

The analysis of role-to-role interaction pairs reveals that within-group persuasion (smallholder to smallholder) was consistently the most frequent and effective channel, accounting for 147 of 187 total opinion changes in ST02 (79\%) and 225 of 291 in ST09 (77\%). Cross-group persuasion from technified producers to smallholders was the second most significant pathway, contributing 17 changes in ST02 and 42 in ST09. This pattern reflects the model's network structure, in which interaction probability is proportional to group size and relational proximity, thereby concentrating discursive exchanges within the numerically dominant smallholder population.

\section{Discussion}

\subsection{Convergence as Discursive Closure}

The convergence toward opposition observed across all four scenarios warrants careful interpretation. Rather than attributing this accelerated convergence solely to the discursive weight of arguments advanced by institutional actors, it also comes from pre-existing distrust that community members hold toward water governance organizations. Empirical survey data from the Longaví River sub-basin ($n = 293$) reveal that 57\% of respondents rated the JVRL as ``not at all present'' (\textit{nada acompa\~nado}) in their territory, with only 13\% perceiving it as ``very present'' \citep{Salgado2025}. This institutional absence is even more pronounced for the Direcci\'on de Obras Hidr\'aulicas (DOH), which 84\% of respondents perceived as absent, and the Dirección General de Aguas (DGA), rated absent by 69\%. Although 78\% of respondents acknowledged that participation in the JVRL brings benefits, 12\% explicitly disagreed, and 60\% of reported water conflicts remained unresolved, with only 18\% of affected users having sought institutional support \citep{Salgado2025}. Thus, it is possible to state that convergence is a form of discursive closure \citep{Hajer1995} in which the accumulation of equity, skepticism, and autonomy-based arguments progressively narrows the argumentative space available for pro-reservoir positions. Discursive closure is consistent with the results from the political ecology of water conflicts in Chile, as repeated institutional failures and perceived distributive injustices have consolidated opposition to large-scale infrastructure among smallholder communities \citep{bauer2015water, budds2013water, Salgado2025}.

The transient support peak observed in ST02 (52\% at $t = 2$ months) before the decline is theoretically significant. It indicates that under low interaction density and without governance intermediaries, initial supporters can briefly consolidate favorable positions before opposition arguments diffuse through the network. This pattern resonates with findings on opinion cascades in agent-based models \citep{Sousa2024, Blair2022}, where early movers can temporarily dominate when connectivity is sparse, but lose influence as information flow increases. The absence of such peaks in TT scenarios confirms that institutional actors altered the discursive landscape from the outset, preventing the formation of transient pro-reservoir coalitions.

\subsection{The Role of Institutional Actors in Discursive Dynamics}

The inclusion of the JVRL and an NGO (TT scenarios) accelerated convergence (7--8 vs. 10--14 steps) and expanded semantic diversity. However, effects are not reducible to the addition of two opposing voices but reflect a qualitative transformation of the argumentative structure. The NGO introduced framing categories (AMB), negligible in the ST scenarios. At the same time, the JVRL's presence, despite its null persuasion rate, altered the institutional context within which other agents formulated their arguments.

Results align with the hydrosocial territories literature \citep{Boelens2014, Hommes2023Need}: governance intermediaries shape deliberative outcomes through direct advocacy and their constitutive role in defining the legitimate terms of debate. Thus, the JVRL's failure to persuade individual agents does not diminish its structural function, but activates institutional references and accountability frames that redirect the discursive repertoire toward equity and governance.

The DOH's consistent inability to change positions (0--10\% success rate) shows what \citet{Zwarteveen2017} and \citet{Swyngedouw2009} describe as the legitimacy deficit of technocratic water governance. The DOH's arguments, framed around aggregate efficiency, storage capacity, and multi-purpose optimization, failed to engage with the distributive, experiential, and trust-based concerns that structured the deliberative field. The survey and interview data \citep{Salgado2025}, which documented low trust in the DOH among water users of the Longaví River sub-basin, support the observed missconnection.

\subsection{Semantic Structuring of Hydrosocial Debate}

The dominance of EQD (21--24\%) and SEG (21--22\%) across all scenarios mimics the empirical results in the Longaví River sub-basin: distributive concerns and water security anxieties are the core interpretive frames through which water users evaluate infrastructure proposals \citep{Salgado2025}. The stability of these categories across different interaction intensities and institutional configurations suggests that they function as what \citet{Strang2023} terms ``deep discursive structures'' (interpretive orientations rooted in material conditions and historical experience that resist short-term discursive shifts).

The rise of skepticism (ESC, 12--16\%) across all scenarios is analytically significant. Rather than representing mere noise or indecision, ESC arguments articulate concerns about institutional trustworthiness, transparency failures, and the gap between policy promises and historical outcomes. The high ESC frequency in TT09 (15.6\%) suggests a paradox: institutional inclusion under high interaction intensity amplifies distrust rather than builds consensus. This result aligns with the ``participation paradox'' in water governance, i.e., increased engagement can expose rather than resolve underlying legitimacy deficits \citep{Zeitoun2020, zeitoun2020analysis}.

Shift from position-level persuasion to system-level claims may explain the late-stage emergence of political arguments (POL). Once agents reached positional consensus, their discursive attention turned to structural demands for institutional reform, subsidy redistribution, and regulatory change. This trajectory also appears in contentious politics literature, where collective opposition evolves from the rejection of specific proposals toward the articulation of broader governance alternatives \citep[e.g.][]{blanco2023water}.

Environmental framings (AMB) remained marginal ($<5\%$) even in scenarios that included the NGO, indicating that ecological concerns occupy a subordinated position within the hydrosocial deliberative field of the Longaví River sub-basin. This result is consistent with empirical observations that environmental arguments, while present in public discourse, are typically subordinated to livelihood-based and distributional claims in smallholder-dominated contexts \citep{budds2004power, Figueroa2020}.

\subsection{Interaction Structure and Persuasion Asymmetries}
The dominance of within-group persuasion (smallholder-to-smallholder 
accounts for 77--79\% of opinion changes) reveals a structural feature 
of the simulated hydrosocial field: discursive influence flows primarily 
along lines of social proximity and shared vulnerability rather than 
across institutional or hierarchical boundaries. This finding aligns 
with network-based models of opinion formation \citep{Alam2022Understanding} 
and with ethnographic accounts of water user communities in Chile, where 
peer-to-peer communication among smallholders serves as the primary channel 
for information and influence \citep{Salgado2025}.

The high per-interaction persuasion rate of the NGO (up to 
67\% in TT02, 25\% in TT09) contrasts with its limited interaction 
frequency, suggesting that certain institutional actors possess concentrated 
discursive efficacy, understood as the capacity to shift positions through 
targeted, high-impact argumentative interventions, rather than through broad 
network presence. Indeed, literature on ``discursive power'' in water 
politics \citep{Boelens2014, De2024Water} states that specific actors 
disproportionately shape outcomes through the quality and framing of their 
contributions rather than through the volume of their participation.

The asymmetry between interaction frequency and influence echoes broader 
debates on participatory governance, where numerical representation does 
not necessarily translate into deliberative impact \citep{Zwarteveen2021}. 
In the simulated system, the DOH's institutional authority did not translate 
into persuasive capacity, while the NGO's limited interactions produced 
disproportionate shifts in position. However, the JVRL's null persuasion 
rate complicates a straightforward reading: unlike the DOH, the JVRL 
combines technical legitimacy with a legal mandate and direct user 
representation, yet equally failed to shift agent positions. This suggests 
that discursive influence in hydrosocial deliberations depends less on the 
formal/epistemic authority distinction than on whether an actor's framing 
resonates with the pre-existing trust dispositions of receiving agents, a 
condition that neither institutional authority nor technical legitimacy can 
substitute \citep{Boelens2014, De2024Water}. Nevertheless, these results 
do not account for persuasion mechanisms beyond direct argumentative 
exchange, such as media control, political influence, or coercive pressure. 
The incorporation of such mechanisms could substantially alter the observed 
patterns, potentially amplifying the discursive efficacy of actors with 
limited numerical representation but greater capacity to exercise structural 
power within the basin.

Beyond these constraints, hybrid architectures of this kind hold 
considerable promise for participatory water governance, less as 
predictive instruments than as exploratory platforms for examining 
how institutional configurations, trust conditions, and discursive 
structures shape deliberation before it formally begins. By making explicit the argumentative repertoires and influence asymmetries that tend to remain tacit in real governance settings, HSABM--LLM-type models could inform the 
design of more contextually grounded consultation processes, helping 
practitioners anticipate discursive bottlenecks and identify the 
conditions under which institutional inclusion is likely to build, or deepen, existing legitimacy deficits.

\subsection{Methodological Implications and Limitations}

The HSABM--LLM framework demonstrates that coupling agent-based models with language models allows for the emergence of discursive dynamics that go beyond rule-based opinion change mechanisms. The model generated arguments in natural language that reflected agent profiles, adapted to contextual summaries, and responded to received messages. Such capabilities are not available in conventional ABMs relying on numerical opinion representations \citep{Taberna2020, Lillo2024Socio}. The semantic classification into nine categories is consistent with the empirical taxonomy, so the LLM maintained thematic coherence throughout simulations.

We acknowledge some limitations in our approach. First, the universal convergence toward opposition across all scenarios may reflect a systematic bias in the LLM toward opposition arguments. The large language model's training corpus likely over-represents critical and skeptical framings of infrastructure projects, which could amplify opposition rhetoric beyond levels observed in empirical deliberation \citep{Ghaffarzadegan_2024, dejeu2024}. Future work should evaluate potential bias through controlled comparisons with alternative LLMs and calibration against observed opinion distributions.

Second, the water supply conditions were held constant across all scenarios. In real hydrosocial systems, drought, water allocation conflicts, or infrastructure failures can abruptly alter the discursive landscape, potentially favoring support for infrastructure \citep{WMO2024}. Integrating a dynamic hydrological component into the ABM would enable the simulation of climate-mediated discursive shifts, thus allowing for the assessment of climate adaptation governance research.

Third, the single-LLM architecture (Qwen3-8B) constrains the diversity of reasoning styles across agent types. While the prompt-based differentiation produced measurable behavioral heterogeneity, ensemble architectures incorporating multiple language models \citep{manduchi2025, lu2024llms} could improve the representation of cognitive diversity across institutional, cultural, and educational backgrounds.

Fourth, the simulated network topology, while grounded in empirical fieldwork, does not capture the complexity of communication channels in the Longaví River sub-basin, including media exposure, informal gatherings, and social media interactions, which could introduce additional influence pathways not represented in the current model.

Fifth, the results are inherently situated: they reflect the 
specific empirical, institutional, and hydrosocial characteristics 
of the Longaví River sub-basin, which served as the design context 
for the experiment. Accordingly, replicability in other settings 
requires contextual adaptation, including adjustments to agent 
typologies, network topology, and the discursive categories 
structuring deliberation. Transferability should therefore be 
understood not as direct model portability, but as methodological 
transposability, contingent on grounding the 
framework in the empirical specificities of each target system.

Despite these limitations, the HSABM--LLM architecture provides a novel platform for exploring how discursive dynamics, institutional configurations, and power asymmetries interact to shape hydrosocial governance outcomes, 
contributing to the integration of political ecology insights into computational simulation frameworks.

\section{Conclusions}

This study demonstrates that coupling agent-based models with large language models enables the simulation of discursive dynamics in water governance at a level of semantic richness beyond that of conventional rule-based approaches. Opinion convergence operates through discursive closure rather than superior argumentation; governance actors accelerate convergence by activating pre-existing distrust rather than persuading. Second, deliberative fields in smallholder-dominated contexts follow a narrow set of deep interpretive frames, principally distributive equity and water security, that resist parametric and institutional variation. Third, formal authority and deliberative influence are decoupled, as institutional presence does not ensure persuasion and, under eroded trust, may deepen opposition. 

Our results suggest that participatory governance design should account for the discursive conditions under which institutional actors enter deliberative spaces, rather than assuming that inclusion is inherently beneficial. 

Future extensions should address potential LLM biases toward opposition framings, incorporate dynamic hydrological feedbacks, and explore multi-LLM architectures to improve the representation of cognitive diversity across actor types.

\appendix

\section{Formal Definitions of Aggregate Indicators}
\label{app:indicators}

This appendix provides the formal mathematical definitions of the five aggregate indicators introduced in Section~\ref{subsec:indicators}. All indicators are computed from the simulation outputs and are used in the Results section to compare the four experimental scenarios.

\subsection*{A.1 Support level}

The \textit{support level}, $\text{Support}^{(t)} \in [0,\; 100]\;\%$, is the proportion of agents expressing support for the reservoir proposal at time step $t$, where $\mathbf{1}[\cdot]$ denotes the indicator function and $o_i^{(t)} \in \{-1, 0, 1\}$ is the opinion state of agent $i$:
\begin{equation}
\text{Support}^{(t)} = \frac{1}{N}
\sum_{i=1}^{N} \mathbf{1}\bigl[o_i^{(t)} = 1\bigr]
\times 100.
\label{eq:support}
\end{equation}

\subsection*{A.2 Discursive diversity}

\textit{Discursive diversity}, $D^{(t)}$, is defined as the complement of the Simpson concentration index \citep{Simpson1949} computed over the relative frequencies of the $K = 9$ semantic categories listed in Table~\ref{tab:categorias_semanticas}. The index ranges from $D = 0$ (all arguments concentrated in a single category) to $D = (1 - 1/K) \times 100 \approx 88.9$ (uniform distribution across all nine categories):
\begin{equation}
D^{(t)} = \left(1 - \sum_{k=1}^{K} p_k^{(t)\,2}\right)
\times 100,
\label{eq:simpson}
\end{equation}
where
\begin{equation}
p_k^{(t)} = \frac{n_k^{(t)}}{\sum_{k'=1}^{K} n_{k'}^{(t)}}
\label{eq:pk}
\end{equation}
and $n_k^{(t)}$ is the number of arguments classified into category $c_k$ at step $t$.

\subsection*{A.3 Opinion change rate}

The \textit{opinion change rate}, $\text{OCR}^{(t)} \in [0,\; 100]\;\%$, is the proportion of dyadic interactions producing an opinion shift at time step $t$. The aggregate rate reported in Table~\ref{tab:resumen_resultados} is computed over all interactions across the full simulation horizon:
\begin{equation}
\text{OCR}^{(t)} = \frac{\bigl|\{(i,j) \in E^{(t)} :
\Delta o(r_{j \to i}^{(t)}) \neq 0\}\bigr|}
{\bigl|E^{(t)}\bigr|} \times 100.
\label{eq:ocr}
\end{equation}

\subsection*{A.4 Persuasion success rate by role}

The \textit{persuasion success rate by role}, $\text{PSR}_r \in [0,\; 100]\;\%$, is the proportion of interactions initiated by agents of role $r$ that produce an opinion change in the receiver, aggregated across the full simulation horizon $T$:
\begin{equation}
\text{PSR}_r = \frac{\sum_{t=1}^{T}
\bigl|\{(i,j) \in E^{(t)} : \text{role}(i) = r,\;
\Delta o \neq 0\}\bigr|}
{\sum_{t=1}^{T} \bigl|\{(i,j) \in E^{(t)} :
\text{role}(i) = r\}\bigr|} \times 100.
\label{eq:psr}
\end{equation}

\subsection*{A.5 Semantic category frequency}

The \textit{semantic category frequency}, $f_k \in [0,\; 100]\;\%$ with $\sum_{k=1}^{K} f_k = 100$, is the relative frequency of semantic category $c_k$ across the full simulation:
\begin{equation}
f_k = \frac{\sum_{t=1}^{T} n_k^{(t)}}
{\sum_{t=1}^{T} \sum_{k'=1}^{K} n_{k'}^{(t)}} \times 100.
\label{eq:freq_semantica}
\end{equation}

\bibliographystyle{elsarticle-harv}
\bibliography{references}

\end{document}